\documentclass{nature}

\usepackage{astjnlabbrev-nature}
\newcommand{\micron}{\mbox{$\mu$m}}
\usepackage{graphicx}

\title{A Spectrum of an Extrasolar Planet}

\author{L.\ Jeremy Richardson$^{1}$, Drake Deming$^2$, Karen Horning$^3$, Sara Seager$^{4,5}$ \& Joseph Harrington$^6$}

\begin{document}

\maketitle

\begin{affiliations}
 \item Exoplanets and Stellar Astrophysics Laboratory, NASA Goddard Space Flight Center, Mail Code 667, Greenbelt, MD 20771
 \item Planetary Systems Laboratory, NASA Goddard Space Flight Center, Mail Code 693, Greenbelt, MD 20771
 \item Department of Physics and Space Sciences, Florida Institute of Technology, 150 W.\ University Blvd., Melbourne, FL 32901
 \item Department of Terrestrial Magnetism, Carnegie Institution of Washington, 5241 Broad Branch Rd., NW, Washington, DC 20015
 \item Department of Earth, Atmospheric, and Planetary Sciences, Massachusetts Institute of Technology, 77 Massachusetts Ave., Cambridge, MA 02139
 \item Department of Physics, University of Central Florida, Orlando, FL 32816
\end{affiliations}

\begin{abstract}
 
Of the over 200 known extrasolar planets, 14 exhibit transits in front of their parent stars as seen from Earth.  Spectroscopic observations of the transiting planets can probe the physical conditions of their atmospheres.\cite{charbonneau02,deming05a} One such technique\cite{richardson03a,richardson03b} can be used to derive the planetary spectrum by subtracting the stellar spectrum measured during eclipse (planet hidden behind star) from the combined-light spectrum measured outside eclipse (star + planet).  Although several attempts have been made from Earth-based observatories, no spectrum has yet been measured for any of the established extrasolar planets.  Here we report a measurement of the infrared spectrum (7.5--13.2~\micron) of the transiting extrasolar planet HD\,209458\,b.  Our observations reveal a hot thermal continuum for the planetary spectrum, with approximately constant ratio to the stellar flux over this wavelength range.  Superposed on this continuum is a broad emission peak centered near 9.65~\micron\ that we attribute to emission by silicate clouds.  We also find a narrow, unidentified emission feature at 7.78~\micron.  Models of these ``hot Jupiter''\cite{colliercameron02} planets predict a flux peak\cite{marley06,burrows05a,sudarsky03,seager98} near 10~\micron, where thermal emission from the deep atmosphere emerges relatively unimpeded by water absorption, but models dominated by water fit the observed spectrum poorly.  
\end{abstract}
\vspace{6pt}


We observed the HD\,209458\,b system during two predicted secondary eclipse events, on 6 and 13 July 2005.  For each event, we observed continuously for 6 hours, centered on the three-hour duration of the eclipse.  We used the InfraRed Spectrograph (IRS)\cite{houck04} on the Spitzer Space Telescope\cite{werner04} in staring mode with the SL1 slit (short wavelength, low resolution), which gives a wavelength coverage of $\sim$7.4--14.5~\micron\ and a spectral resolution ($\lambda/\Delta\lambda$) of 60--120.  We analyzed a total of 560 individual spectra of the system (280 per eclipse event), each with integration time 60.95~sec, in order to obtain a single spectrum of the planet for each event.


Our technique\cite{richardson03a,richardson03b} exploits the timing of the eclipse to derive the planetary spectrum from the eclipse depth vs.\ wavelength.  Our analysis effectively uses IRS as a multi-channel photometer by searching for the eclipse in the time series of flux at each wavelength.  This method is equivalent to subtracting the in-eclipse spectra (planet hidden) from the out-of-eclipse spectra (both star and planet visible).  We developed a custom procedure to extract flux spectra from the IRS images, and we verified that our results are robust with respect to the details of this spectral extraction.  The Supplementary Information (SI) presents a complete discussion of our methodology.

We first verify that the eclipse is visible in the wavelength-integrated flux, as shown in Figure~\ref{fig:phot}.  The eclipse is clearly visible in the total flux, appearing as a dip centered on phase 0.5.  The depth of the eclipse is not easily determined from this plot alone, due to three systematic effects;
these are noted in Figure~\ref{fig:phot} and their corrections are discussed in the SI.  

To derive the planetary spectrum, we use a differential analysis. Recasting the 280 spectra per eclipse as flux vs.\ time at each wavelength, we divide by the average spectrum and subtract the average time series to produce residual fluxes.  This subtracts two of the systematic effects (the baseline ramp and the telescope pointing oscillation; see Figure~\ref{fig:phot} and the SI).  We then fit a model eclipse curve to the time series of residual fluxes at each wavelength; the amplitudes from these fits  comprise the planetary spectrum.  The remaining systematic effect, a slow drift in telescope pointing, is corrected by our calibration procedure, placing the spectrum in contrast units (ratio of planetary flux to stellar flux).

The upper panel of Figure~\ref{fig:spec} shows the derived planetary spectra from both eclipse events separately, which allows us to confirm the reality of spectral structure.   Two real spectral features are present above the noise level and are seen in both eclipse events:  1) a broad feature centered near 9.65~\micron; and 2) a sharp feature occupying only a few wavelength channels centered near 7.78~\micron\ (which is confirmed by a separate analysis shown in Figure~\ref{fig:hipass} and detailed in the SI).  Both of these spectral features appear in emission, i.e., in excess of the apparent continuum level. 
The middle panel (Figure~\ref{fig:spec}b) shows the average of the two eclipse events.  A chi-squared analysis confirms the presence of structure in the spectrum.  Specifically, a flat line (i.e., constant contrast) is inconsistent with the data at the 3.5$\sigma$ level.  Recall that the eclipse is seen clearly in Figure~\ref{fig:phot}.  After correcting for systematic errors, the eclipse depth ($\sim0.3$\%) exceeds the errors at individual wavelengths in Figure~\ref{fig:spec}.  
Therefore, a flat line in Figure~\ref{fig:spec} would also represent a clearly detected, but structureless, spectrum.  The reality of the broad feature at 9.65~\micron\ is further illustrated by the lower panel of Figure~\ref{fig:spec}.  This plot shows the average spectrum from the middle panel, binned coarsely over wavelength.  The rise in flux in the region between 9 and 10~\micron\ is clear and statistically significant ($3.6\sigma$ difference between flux points at 9 and 10~\micron).  Several other suggestive features are apparent in Figure~\ref{fig:spec}a (e.g., possible absorption at 8.6 and 9.3~\micron), but these are not clearly detected.


We now consider interpretations of the two features observed in the measured spectrum (and summarized in Table~\ref{tbl:res}).  First, the 9.65-\micron\ peak (most noticeable as a rise in the spectrum from 9--10~\micron, as shown in Figure~\ref{fig:spec}c) is significant at the $3.6\sigma$ level when suitably binned to the apparent width of the feature.  Because of this peak and the relatively flat spectrum at 10--13~\micron, blackbody spectra (in the temperature range 1100--1600~K) are ruled out to the $\sim 3.5\sigma$ level.  A seemingly natural interpretation of this feature is water vapor absorption at 7--9~\micron.   Such an absorption feature is prominent in most published HD\,209458\,b models.\cite{marley06}  All hot Jupiter spectra are expected to be shaped by water absorption because water is an abundant gas at the high temperatures of hot Jupiters (1000--2000~K).  However, we do not favor this water absorption interpretation. We previously reported an upper limit on the water vapor absorption feature at 2.2~\micron\ for the spectrum of this planet.\cite{richardson03b}   Moreover, based on Figure~\ref{fig:spec} alone, a typical solar abundance model of HD\,209458\,b with strong water features\cite{seager05} is ruled out at the 3.5$\sigma$ level, due to the poor fit to the spectral slope at the shortest wavelengths.  
Our results for the contrast in this spectral region are consistent with the depth of the secondary eclipse at much longer wavelength (24~\micron).\cite{deming05b}

The occurrence of a peak at 9.65~\micron\ is strongly reminiscent of the Si-O fundamental stretching mode at 9.7~\micron,\cite{dorschner71} manifested in this case as silicate clouds.  Absorption and emission from amorphous and crystalline silicates are ubiquitous in young star- and planet-forming regions,\cite{kessler-silacci06} and silicates can also condense directly in hot Jupiter atmospheres.\cite{burrows99,seager00b}  
Recent observations\cite{cushing06} of L dwarfs reveal 10~\micron\ absorption by silicate clouds.  The silicate grains must be small ($<10$~\micron) to exhibit the feature,\cite{cushing06} suggesting that they can occur at high altitudes.  Further, to produce a silicate feature in emission requires that silicate clouds be present in a region of inverted temperature gradient.  We hypothesize that the feature could be explained by high silicate clouds in the stratosphere with an inverted temperature gradient.
Several recent studies have suggested the possibility of a deep stratosphere on hot Jupiters.  The discovery of OGLE-TR-56b prompted models that include strong stellar irradiation, and one study concluded that TiO in the upper atmosphere can cause a temperature inversion.\cite{hubeny03} 
More recently, the detection of thermal emission from TrES-1\cite{charbonneau05} using IRAC revealed a higher brightness temperature at 8~\micron\ than at 4.5~\micron, which was unexpected based on previous models, and one of several explanations is a temperature inversion.\cite{fortney05a,fortney06a}  
Finally, in this respect, we note that the presence of high clouds (to $\sim$millibar pressures) is consistent with other observational results for this planet, specifically the low sodium abundance,\cite{charbonneau02} the upper limit on CO absorption during transit,\cite{deming05a} and the non-detection of water bands in the near-IR.\cite{richardson03b}  Unanticipated sources of opacity may be required to produce a temperature inversion at these altitudes, and thereby mask the effect of water opacity.

Alternatively, the planet is known to have an extended and evaporating atmosphere,\cite{vidal-madjar03} and it is possible that an optically-thin, emitting dust envelope could contribute to the 9.65~\micron\ feature.  We also caution that our silicate feature is based on a rise in the spectrum near the Si-O stretching resonance, and at the level of data uncertainty we do not claim a down turn beyond 10~\micron\ that  would support the silicate feature claim.

The second feature in our spectrum is a narrow, sharp peak at 7.78~\micron.  This peak is statistically significant at the 4.4$\sigma$ level and is unlikely to be an instrumental error because the peak appears in the spectra from both observed eclipse events.  If produced by thermal emission, this feature is also consistent with an inverted temperature gradient.  We considered the possibility that this peak is due to methane emission.  Figure~\ref{fig:hipass}a includes a 
profile of the wavelength dependence of methane emission, obtained by scaling the HITRAN\cite{rothman05} line strengths to T=1500K, binning them to IRS resolution, and assuming optically-thin emission.  The observed peak is not coincident with the strongest methane lines (Q~branch). The predicted position of the Q~branch shifts to longer wavelength with increasing temperature,  but a two-pixel discepancy remains at any plausible temperature, and a wavelength calibration error of this magnitude is out of the question (Spitzer Support, private communication).  Although  other methane features occur over the range extending from $\sim$7.4 to $\sim$8.0~\micron,\cite{rothman05} it seems unlikely that these weaker lines alone could cause the feature in the observed spectrum.  A more exotic possibility that cannot be firmly rejected is the C-C stretching resonance in polycyclic aromatic hydrocarbons.\cite{sloan05}
Additional Spitzer observations should clarify the nature of this emission.

Finally, we look forward to future extension of extrasolar planet spectroscopy to the domain of transiting terrestrial planets.  Although Spitzer's modest size currently limits us to the brightest transiting planet systems,  the 6.5-m aperture of the forthcoming James Webb Space Telescope\cite{gardner06} should provide a sufficient photon flux to to measure the spectrum of a transiting ``hot Earth'' orbiting a nearby lower-main-sequence star.\cite{tarter06} 

\vspace{1in}
\bibliography{refs,richardson}

\begin{addendum}
\item[Supplementary Information] accompanies the paper on {\bf www.nature.com}.
 \item 
This work is based on observations made with the Spitzer Space Telescope, which is operated by the Jet Propulsion Laboratory, California Institute of Technology under a contract with NASA. Support for this work was provided by NASA.  We gratefully acknowledge cooperation with Dave Charbonneau, Carl Grillmair, and Heather Knutson.  Our understanding of the long-term telescope pointing drift was derived from Knutson's study of the effect in their 30~hour program to measure the light curve of HD\,189733\,b.  Charbonneau was kind enough to provide his measurement of the eclipse depth of HD\,209485\,b that was key to casting our results in terms of contrast, rather than differential contrast.  We also thank Mark Swain and Amanda Mainzer for discussions regarding the telescope pointing oscillation.
We thank the the teams that designed, built, operate, and support the Spitzer Space Telescope and the IRS.  
We also thank the NASA Astrobiology Institute, which has centers at both NASA Goddard and the Carnegie Institution of Washington. 
LJR acknowledges support as a NASA Postdoctoral Fellow at NASA Goddard (formerly the NRC Research Associateship Program).  KH performed the high-pass filtering analysis as part of her participation in the Summer Undergraduate Internship in Astrobiology, funded by the Goddard Center for Astrobiology.  SS thanks the Spitzer Theory Program and the Carnegie Institution of Washington for support.  
 \item[Competing Interests] The authors declare that they have no
competing financial interests.
 \item[Correspondence] Correspondence and requests for materials
should be addressed to LJR\\
(email: lee.richardson@colorado.edu).
\end{addendum}

\clearpage



\begin{figure}
\begin{center}
\includegraphics[scale=0.75]{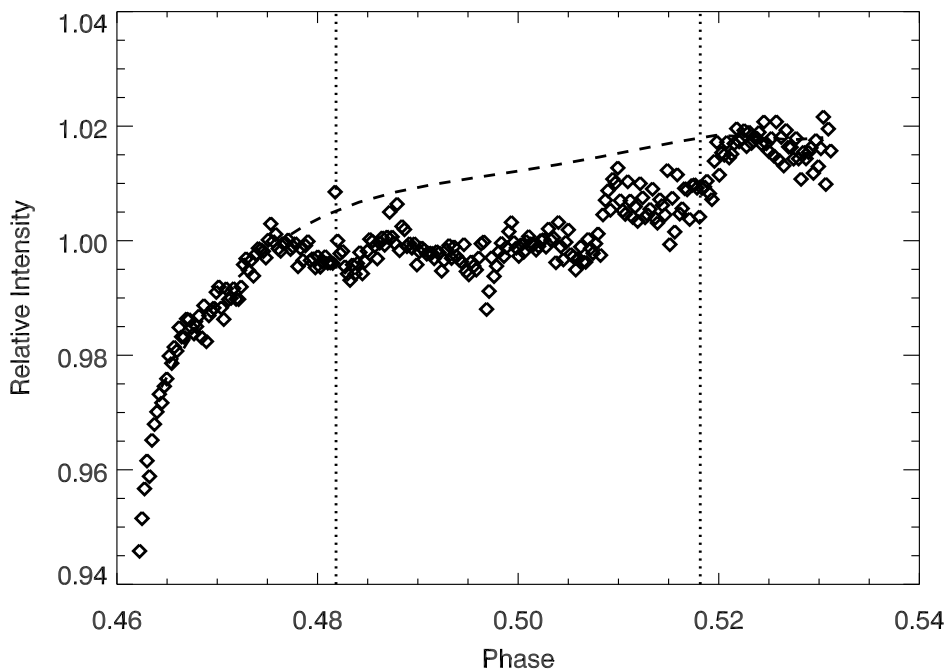}
\caption{Wavelength-integrated flux as a function of orbital phase, showing the detection of the secondary eclipse, centered at phase 0.5.  The plot shows the total flux, calculated by summing all wavelengths for each spectrum.  The results for the two eclipse events are then added together and normalized to the mean value of the total flux of both events.  The eclipse (with apparent depth $\sim$0.5\%) is observable in spite of several systematic effects.  The known boundaries of the eclipse (first and fourth contacts) as derived from data in the visible\cite{brown01a,knutson07} are indicated by the vertical dotted lines.
Three systematic effects (see the SI for details) are removed by our analysis and are present in this figure: 1) a slow ramp-up in intensity of the baseline\cite{deming06} (dashed curve); 2) a telescope pointing oscillation of 1.02-hour period that modulates the flux transmitted through the instrument slit (although this effect is difficult to see here, since the oscillation was nearly out of phase for the two eclipses); and 3) a slow drift in telescope pointing that causes an extra dip in intensity and adds to the apparent depth of the eclipse.
  \label{fig:phot} }
\end{center}
\end{figure}

\begin{figure}[t!]
\begin{center}
\includegraphics[scale=0.45]{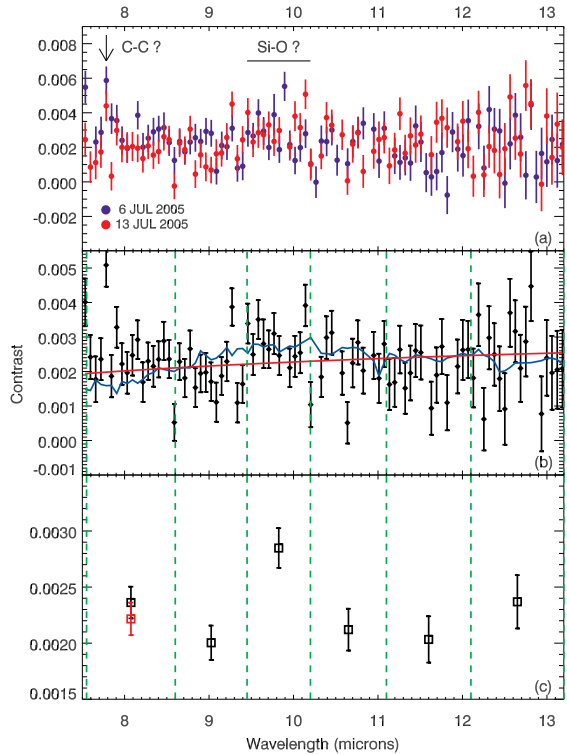}
\caption{The spectrum of HD\,209458\,b from 7.5--13.2~\micron. The upper panel (a) indicates the result from both observed eclipse events separately.  Emission features (near 7.8~\micron\ and 9.65~\micron) and candidate identifications are indicated.
The units of the y-axis are contrast (planet relative to star), and the absolute depth of the eclipse has been calibrated to the preliminary IRAC result at 8~\micron\ (Charbonneau, private communication); see the SI for details.
We make no claims about the spectrum beyond $\sim$10.5~\micron, where the errors increase due to the decreasing flux and the points are correspondingly more scattered.  The middle panel (b) shows the average of the two events with models overplotted.  The blue curve is a model for HD\,209458\,b\cite{seager05} (which is consistent with the photometric result at 24~\micron\cite{deming05b}); the red curve is a 1600~K blackbody for the planet divided by a 6000~K blackbody for the star (although a range of blackbody temperatures for the planet from 1100--1600~K were tested).  The lower panel (c) shows a coarse binning in wavelength (boundaries indicated by green dashed lines) of the average spectrum from panel (b).  The bins were defined to probe the spectral features we discussed.  The weighted mean of all the points in each bin is calculated, and the error on the mean is also weighted by the errors on the individual points.  An average of 14~data points appear in each bin.
For the bin at the shortest wavelength, two points are shown: one including the 7.8~\micron\ emission feature (black) and one excluding this feature (red).
For all three panels, the error bars represent $\pm$ s.e.m; i.e., they are calculated by propagating the errors in the individual points to determine the error on the mean.
Also for all three panels, we show only the result shortward of 13.2~\micron\ because IRS spectra at the longest wavelengths are affected by a systematic error called the ``teardrop effect.''  This effect is not well understood but is believed to be caused by scattered light (see the IRS Data Handbook v.2.0, p. 62).
  \label{fig:spec} }
\end{center}
\end{figure}

\begin{figure}
\begin{center}
\includegraphics[scale=0.75]{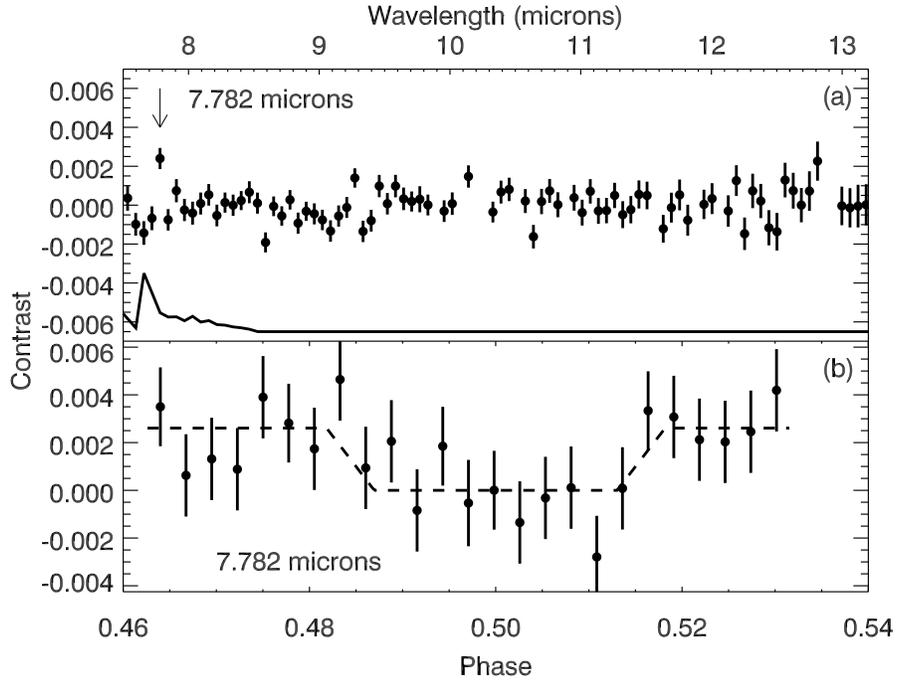}
\caption{Separate analysis to confirm the unidentified emission feature near 7.8~\micron.  Since the primary instrument systematics vary slowly with wavelength, an alternative method to eliminate them is to apply a high-pass filter to the observed spectra (which also suppresses any broad planetary spectral structure).  We fit a sixth-order polynomial to each spectrum and subtract this fit.  Forming a time series of the filtered differences at each wavelength, we fit the model eclipse curve as before.  The resulting strongly-filtered planetary spectrum, averaged for the two events, is shown in the upper panel (a).   The solid line is a model of methane line strengths from HITRAN\cite{rothman05} scaled to 1500~K; units are arbitrary and illustrate relative line strengths.
The 7.78~\micron\ point stands out, as shown in the upper panel.
The point is detected at the 5.4$\sigma$ significance level, calculated in the same way as described in Table~1 but using the high-pass filtered spectra; the significance level is higher because the wings of the feature have been supressed.
The lower panel (b) shows the binned time series for the single wavelength channel at 7.78~\micron. The model eclipse fit to this time series is overplotted (dashed line), indicating the differential eclipse is visible at this wavelength with the correct duration and central phase.  The error bars in both panels represent $\pm$ s.e.m., calculated by propagating the errors from the individual points to determine the error on the mean.
\label{fig:hipass} 
}
\end{center}
\end{figure}












\begin{table}
\begin{center}
\caption{Detected features in the spectrum of HD\,209458\,b.
For the narrow feature at 7.8~\micron, the width is estimated by fitting the shape of the feature, as described in the SI.
The average contrast is computed by the taking the single average point at 7.8~\micron\ and subtracting the mean of the 2 pixels on both sides of the peak (4 pixels total).  For the broad feature, the width of the feature is only a rough estimate, since we do not claim that the feature has a definite downturn beyond 10.5~\micron.  The average contrast and error are based on the binned spectrum in Figure~\ref{fig:spec}c.  Here we take the ``peak'' bin near 10~\micron\ and subtract the ``continuum" bin at 9~\micron\ to get the average contrast, and the error is the relative error between the two points.  For both features, the significance level is simply the average contrast divided by the s.e.m.
	\label{tbl:res} }
\vspace{0.2in}
\begin{tabular}{ccccccc}
$\lambda_{min}$ & $\lambda_{max}$ & Number of & Average & Standard & Significance & Candidate\\
(\micron) & (\micron) & Channels & Contrast & Error (s.e.m.) & Level ($\sigma$) & Identification\\\hline
$\sim$7.65 & $\sim$7.92 & $\sim$ 2 & 0.0027 & 0.00063 & 4.4 & C-C ?\\
$\sim$9.3 & $\sim$10.1 & $\sim$13 & 0.00085 & 0.00023 & 3.6 & Si-O\\

\end{tabular}

\end{center}
\end{table}


\clearpage

{\Large\bfseries\noindent\sloppy \textsf{Supplemental Information: Methodology} \par}%

\vspace{0.5in}

\begin{abstract}

Here we describe the details of our methods and analysis to derive the
planetary spectrum from our IRS observations of HD\,209458\,b.  The
main result is summarized in Supplemental Figure~\ref{fig:sispec}, which shows the
average spectrum of the planet for the two observed eclipse events
(compare to Figure~2b in main text).  In order to allow other
researchers to compare models to our observed spectrum, we provide the final planetary spectra in contrast units (planetary to stellar flux) for both events as a text file (also available for download as Supplementary Information).

\end{abstract}

\begin{figure}[h!]
\begin{center}
\includegraphics[scale=0.65]{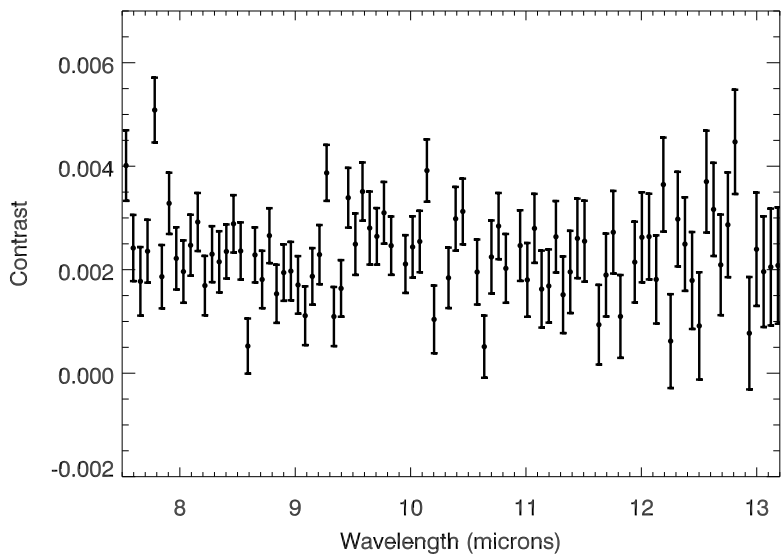}
\caption{Average planetary spectrum from the two eclipse events.
	\label{fig:sispec} }
\end{center}
\end{figure}

\section{Spectral Extraction}
We created a custom procedure for extracting the combined light
spectra (star+planet) from the IRS images.  We read in the Basic
Calibrated Data (BCD) images, created from raw data by the reduction
pipeline at the Spitzer Science Center (version S13.2.0).  Each BCD is
a 2-D image representing a single 60.95-sec integration on the
source. The dimensions of the image represent wavelength and spatial
distance parallel to the spectrograph slit.  The standard procedure
for IRS Staring Mode is to observe half the images with the star at
position A on the slit, nod the telescope, and then record the other
half of the observations at position B.  We found this procedure to be
optimal for the purpose of extracting the planet spectrum.

The 560 BCD images are separated into four groups (of 140 images each) based on eclipse event and nod position.  For each group, we perform the following steps:

\begin{enumerate}
\item {\em Identify and correct bad pixels.}  We employed the {\tt
IRSCLEAN\_MASK} (created by J.\ Ingalls and the IRS Instrument Support
Team) routine downloaded from the Spitzer website.  It accounts for
known bad pixels on the array and allows the user to identify and
correct other ``rogue pixels'' by inspecting the 2-D data.

\item {\em Median filter the images.}  The median image is derived by
calculating the median value of each pixel from the stack of images.
For the A position, we exclude the first 50 images from the median
calculation to avoid bias due to the systematic ramp that occurs at
the beginning of each observing sequence (see Figure~1 in the main
text).  The median image is subtracted from each individual image, and
we then apply the {\tt SIGMA\_FILTER} routine (see the IDL Astronomy
Library) to the difference image; we reject and correct a given pixel
(centered on a box of width 5 pixels) if it exceeds 10$\sigma$ of the
pixels in the box.  This serves to correct any rogue pixels not
identified in Step 1.

\item {\em Create the background-subtracted images} by subtracting
from each individual image the median image from the {\em opposite}
nod position.  For wavelengths near the center of our bandpass, the
background is about 2 percent of the stellar intensity. We checked the
hypothetical possibility that background fluctuations, not removed by
the nod, might affect our results. This was checked two ways; first,
we performed the entire analysis without subtracting the background,
finding essentially the same results, but with higher noise.  Second,
we produced a set of background spectra, by extracting the background
in each image as if the star were present, but using the opposite nod
position from the star.  Analyzing these spectra in lieu of the real
data, we see no effects in the background that would contaminate our
planet spectrum.

\item {\em Extract the spectrum from each image.}  We find the maximum value (peak) in the spectrum at each row (which represents
wavelength), and we add the 4 pixels on either side of the peak,
orthogonal to the dispersion direction (for a total of 9 pixels) to
obtain the flux at this wavelength.  We ignore the curvature of the
spectrum on the array.  We assign the wavelength of each point by
averaging the corresponding pixels in the calibration file {\tt
b0\_wavsamp\_wave.fits} from each event.  Two members of our team
extracted the spectra using separate analysis routines based on this
method, and we ran both sets of spectra through our entire analysis,
obtaining virtually identical results. We also used the SPICE software
from the SSC to extract spectra, and we verified that these spectra
are also consistent with the conclusions of this paper, again by
running them through the entire analysis. However, the SPICE spectra
do not produce as high a signal-to-noise ratio in the planet spectrum
for this very specialized problem.  We also varied the width of the
window to 6 pixels on either side of the peak (13 pixels total), and we ran these spectra through our entire analysis. In deciding which version of the spectra to use for our final analysis, we computed the chi-squared
statistic for the difference between the planet spectra derived for
the two eclipse events, and we use the spectra (9-pixel width, custom extraction) that produce the minimum chi-squared in the difference.

\item {\em Apply a multiplicative factor}.  This factor varies with
wavelength, and it essentially corrects for the discontinuity caused
by the telescope nod and imperfect flat-fielding of the detector
array.  In order to avoid bias in this correction, the factor is
calculated using only the data with the planet in eclipse (i.e.,
behind the star), and separate factors are calculated and applied for
each wavelength.  At this point, the A and B spectra are recombined,
and from here onward, we consider only two groups of spectra,
corresponding to the two eclipse events.  Note, however, that our
analysis is done independently at each wavelength using the time
series of intensity.  Thus, combining the spectra specifically means
that we are adjusting the time series at each wavelength to fix the
discontinuity caused by the nod.

\begin{figure}[t!]
\begin{center}
\includegraphics[scale=0.75]{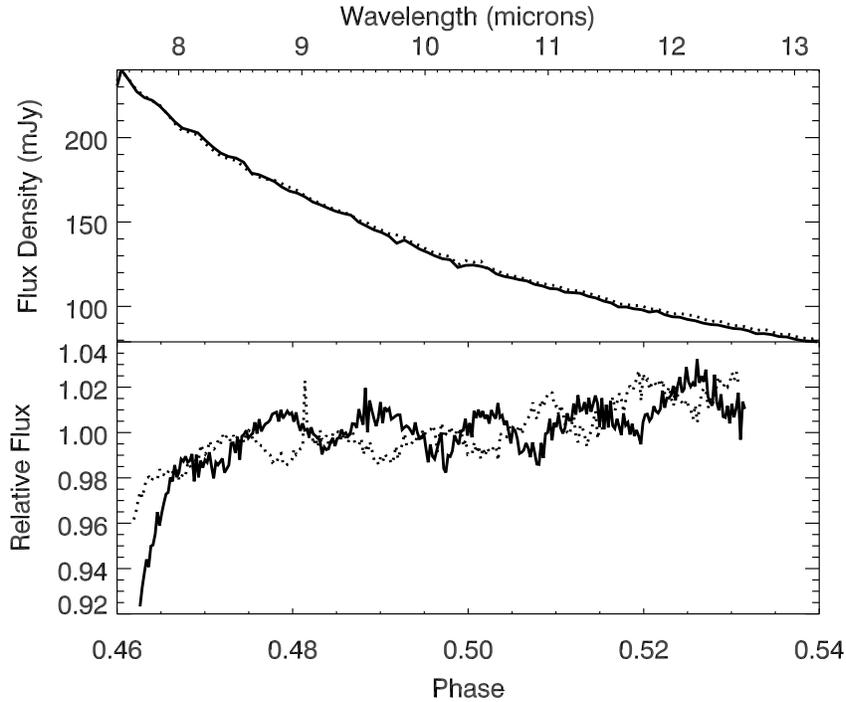}
\caption{Normalizations used in the analysis.  Upper panel shows the
average spectrum for each eclipse event (solid line for first event,
dotted line for second event), and the flux density is approximate because we have not accounted for slit losses.  Lower panel similarly shows the average time series for both events.
	\label{fig:norm} }
\end{center}
\end{figure}

\item {\em Normalize the spectra} by dividing each spectrum by the
average spectrum, shown in Supplemental Figure~\ref{fig:norm}a, in that eclipse
event (to convert the spectra to ``contrast'' units).  This step
essentially just normalizes the intensity in each time series to
unity.  We then subtract the average time series, shown in
Supplemental Figure~\ref{fig:norm}b, from the individual time series at every
wavelength point.  This serves to remove the first two systematic effects in the data (see Section~\ref{sec:sys}).  Supplemental Figure~\ref{fig:stacks} shows the ``stacks'' of spectra (wavelength on the x-axis and phase on the y-axis) before and after the normalization.  

\begin{figure}[t!]
\begin{center}
\includegraphics[scale=0.75]{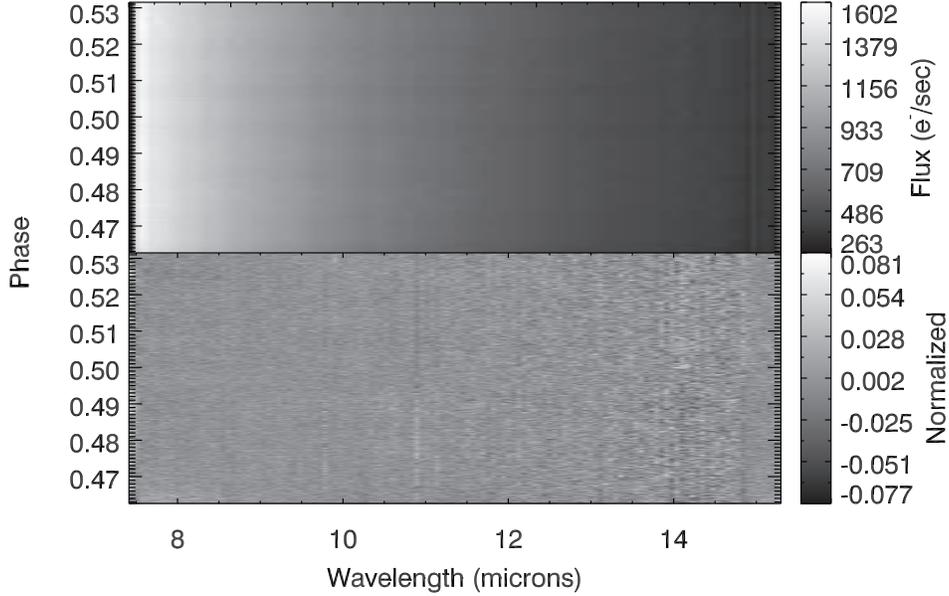}
\caption{Spectra obtained for first eclipse event.  Upper panel shows
the uncorrected spectra (after executing Step 5).  Lower panel shows
the normalized residual spectra (after executing Step 6).
	\label{fig:stacks} }
\end{center}
\end{figure}

\item {\em Calculate the errors} by computing the standard deviation in each time series of normalized contrast values at each
wavelength. The standard deviation is calculated by shifting the time
series by one step, subtracting the shifted series, and computing the
standard deviation of the difference, with the end points omitted. We
divide this precision by the square-root of 2, and we assign it to all
contrast values in the time series at that wavelength.  The error
ranges from 0.0055--0.029 (noise is higher at longer wavelengths).
These values are within 50 to 70\% of the fractional statistical
fluctuations in the number of photons (electrons) detected.  These
per-point precisions are propagated through the linear regressions
(see below), resulting in errors assigned to the planet spectrum at
each wavelength.
\end{enumerate}

\section{Fitting the Eclipse Curve}
\label{sec:fit}

{\em We fit a model eclipse curve to the time series at each
wavelength.} The eclipse amplitude (duration and central phase held
fixed) is estimated using multiple linear regression, simultaneously
with a residual linear ramp and a residual periodic oscillation.  The
periodic oscillation used as the independent variable in the
regression is obtained by subtracting from the average time series a
fourth-order polynomial fit to the average time series; the result is
the oscillation alone with the correct period.  The amplitude of the
eclipse fit at each wavelength gives the planetary spectrum, and is
nearly equivalent to subtracting the in-eclipse combined light
spectrum from the out-of-eclipse combined light spectrum, which we
verified by actually subtracting those spectra.  By allowing for a
residual linear ramp in the linear regressions, we are drawing on our
experience\cite{deming06} that the differences in the ramp from pixel to pixel are linear to a good approximation, except for the first
$\sim$30 integrations, where higher order effects can sometimes
remain. We experimented with omitting the first 30 integrations from
the regressions, but this had little effect on our results.  By
fitting for a residual 1.02~hr oscillation in the regressions, we
allow for the possibility that the oscillation may not be perfectly
subtracted by our procedure.  However, we found that our eclipse
amplitudes are remarkably insensitive to fitting (or not) for a
residual oscillation.  We attribute this to the fact that the time
scale of the oscillation is several times shorter than the eclipse
duration, and this bandwidth difference mitigates significant
interaction with the fitted eclipse amplitude. Errors on the fitted
eclipse amplitudes are returned by the regression routine, based on
the per-point precisions described below.

{\em Reject wavelengths where the reduced chi-squared of the
eclipse fit is greater than a cutoff value.}  We found that the linear
regressions usually fit the contrast time series data to within the
noise.  Accordingly, the distribution of reduced chi-squared values is
centered closely on unity, and we reject any wavelength where the
reduced chi-squared exceeds 1.2.  This stringent criterion eliminates
false-positive detections of residual eclipses, so they are not
manifest as spurious features in the planet spectrum.  Of 94 wavelength bins shortward of 13.2~\micron, we reject 9 bins from the first eclipse event and 8 from the second event.

\section{Calibration and Consistency Checks}
\label{sec:calib}

{\em Calibrate the results to contrast units.} The resulting spectra (one for each eclipse event) are averaged together (Supplemental Figure~\ref{fig:sispec}). We checked to ensure that we obtain the same result regardless of whether the binning or averaging was performed first. Prior to averaging the two eclipses, a calibration is applied to the planet spectrum from each event to adjust for the wavelength variation of the slit losses from slow image drift, and to place the spectra on an absolute contrast scale. The measured preliminary depth of the eclipse (0.25\%) in the IRAC 8~\micron\ band (kindly communicated by D. Charbonneau in advance of publication) is used as a calibration for the effect of slit losses.  We weight the wavelengths in our spectra so as to simulate the IRAC 8-micron bandpass as closely as possible within the limit imposed by the incomplete wavelength overlap. We removed the telescope oscillation using a Fourier notch filter.  This produces a synthetic IRAC eclipse from our IRS data (as in Figure~1 of the main text), and we average it over the two eclipses.  This average eclipse is too deep, due to a component of slow telescope drift perpendicular to the slit.  We scale the extra depth with wavelength, and subtract these corrections from the contrast values in our planet spectrum. The primary effect of this procedure is a zero-point correction to the
contrast; the variations with wavelength are smaller ($<0.1$\%), and
vary gradually over our bandpass.  The scaling with wavelength is
based on the measured wavelength dependence of the intensity
fluctuations created by the 1.02-hour telescope oscillation, since the
wavelength scaling is independent of temporal frequency. 

{\em Apply a high-pass filter analysis,} as an alternate procedure for removing the telescope and instrument systematics to check the reality of the sharp spectral feature at 7.78~\micron.  This procedure begins by fitting each spectrum with a high-order polynominal, and removing this fit to yield residual intensities.  Both fourth and sixth order polynomials were used, and produced similar results. The principle of this analysis is that most telescope and instrument systematics vary slowly with wavelength (see Section~\ref{sec:sys}) and are removed by the polynomial fit.  However, sharp spectral features will remain, and their differential eclipses will be detectable by linear regression, as described above.  As for the main analysis, we rejected poor fits to avoid false-positive detection of residual eclipses; for this analysis we
tightened our limit in reduced chi-squared to 1.15.  Of the 94 wavelength bins shortward of 13.2~\micron, we reject 7 bins from the first eclipse event and 13 bins from the second event.  We find eclipses in the 7.78~\micron\ feature with correct ingress and egress times that repeated for both eclipse events.

\begin{figure}[t!]
\begin{center}
\includegraphics[scale=0.75]{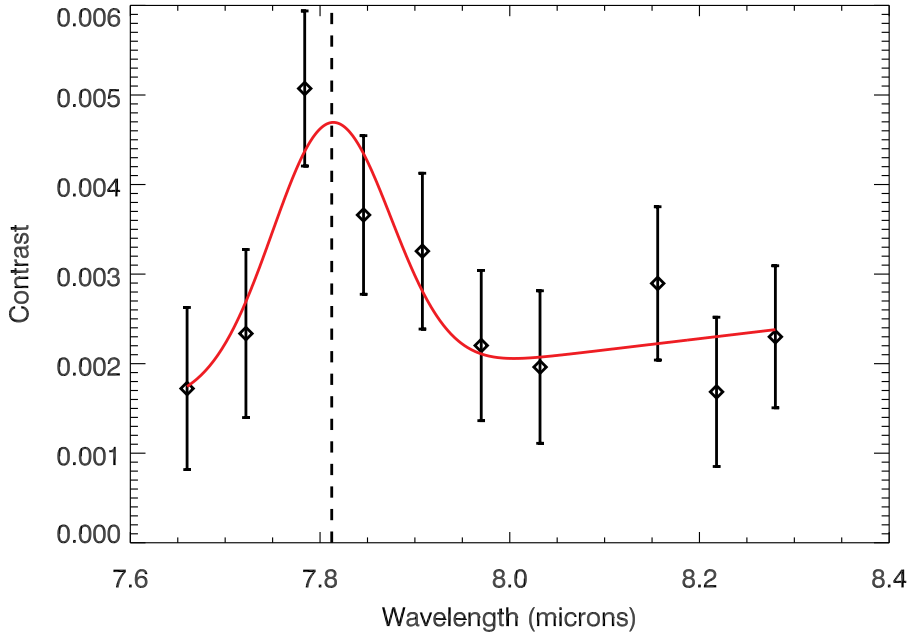}
\caption{Fit of Gaussian plus linear baseline to the unidentified feature near 7.8~\micron.  Data points represent the average of the two eclipse events, with one exception:  the point at 8.25~\micron\ is the value at that wavelength from the first eclipse only, since the point from the second event is clearly discrepant (see Figure~2a of the main text).
	\label{fig:profile} }
\end{center}
\end{figure}

{\em Determine the width of the 7.78~\micron\ feature.} Another key test for the reality of the 7.78~\micron\ feature is to determine whether it is consistent with the 2-pixel spectral resolution of IRS.  A feature occupying a single pixel, for example, is not likely to be real, since even an instrinsically sharp feature will be broadened to the 2-pixel resolution of the instrument.  Therefore, we measured the width of this emission by fitting a Gaussian profile to the data. The
strong-filtering analysis has the property that it suppresses broad
spectral structure such as the 9.65~\micron\ feature, and it also
attenuates the wings of the 7.78~\micron\ feature.  We fit to
the width of the 7.78~\micron\ feature in the data from our primary analysis (Figure 2 of the main text), and this fit is shown in Supplemental Figure~\ref{fig:profile}.  The result indicates a width of  2.2 pixels, in agreement with the spectral resolution of the observations.

\section{Systematic Effects}
\label{sec:sys}
We now discuss the three systematic effects in the data and their removal.
\begin{enumerate}
\item The first systematic effect is a gradual increase in intensity over the 6 hours of each eclipse, that we denote
as the ``ramp''. The ramp\cite{deming05b} is shown in
Supplemental Figure~\ref{fig:norm}b (but is seen more clearly in Figure~1 of the
main text). The cause of the ramp is not completely understood, but is likely related to charge trapping in the Si:As detector material
(H. Knutson and D. Charbonneau, private communication). The other systematic effects are due to telescope pointing errors.  These
pointing errors are comprised of a 1.02-hour oscillation, and a drift
on longer time scales, both described below.  To first order, both the
ramp and the 1.02-hour oscillation are removed by subtracting the
average time series as described above. (This subtraction also removes the average eclipse, but still allows us to derive the planet spectrum by finding the differential eclipse depths as a function of
wavelength.)  

\item The 1.02~hr intensity oscillation is due to a periodic telescope
pointing error, shown in Supplemental Figure~\ref{fig:norm}b. This effect is well known to the SSC, and we used our data to verify that the telescope
pointing oscillation is indeed the cause of the intensity oscillation
seen in our spectra.  We measured the spatial position of the star
along the slit, by fitting to the centroid of the spatial intensity
distribution at each wavelength.  We find that these positions show
the same 1.02~hr oscillation, and are strongly correlated with the
intensity oscillation seen in our data.  We also verified that the
phase of the intensity oscillation is independent of wavelength, and
its amplitude is weakly dependent on wavelength.  The wavelength dependence is measured and found to be consistent with expectations based on diffraction of the PSF.  By subtracting the average time series from the individual time series at each
wavelength, we very effectively remove the oscillation.  However, our analysis also includes residual oscillation that remain at some wavelengths (as mentioned in Section~\ref{sec:fit}).

\item The third systematic effect is a slow telescope drift that
causes the depth of the eclipse (Figure~1 of the main text) to
appear deeper than it actually is.  The existence of this slow drift
was indicated by the position of the star parallel to the slit, that
was derived as part of our analysis.  We have no direct information
on stellar motion perpendicular to the slit for our observations, but
D. Charbonneau and H. Knutson showed us position information from their 30-hour sequence of IRAC photometry on HD\,189733, and this revealed significant drift in both orthogonal coordinates, on long time scales. Hence we developed a calibration procedure that removes the effect of slow drift from our observations in a very general manner, as described in Section~\ref{sec:calib}.
\end{enumerate}

\end{document}